\documentclass[showpacs,prl,superscriptaddress,preprint]{revtex4}
\usepackage{amssymb}
\usepackage{graphicx}
\usepackage{endnotes}

\begin{document}

\title{First Principles Analysis of Electron-Phonon Interactions in Graphene}

\author{K. M. Borysenko\footnotemark[1]\footnotetext[1]{These authors have contributed equally}}
\affiliation{Department of Electrical and Computer Engineering,
North Carolina State University, Raleigh, NC 27695-7911}

\author{J. T. Mullen\footnotemark[1]}
\affiliation{Department of Physics, North Carolina State University,
Raleigh, NC 27695-8202}

\author{E. A. Barry}
\affiliation{Department of Electrical and Computer Engineering,
North Carolina State University, Raleigh, NC 27695-7911}

\author{S. Paul}
\affiliation{Department of Physics, North Carolina State University,
Raleigh, NC 27695-8202}

\author{Y. G. Semenov}
\affiliation{Department of Electrical and Computer Engineering,
North Carolina State University, Raleigh, NC 27695-7911}

\author{J. M. Zavada}
\affiliation{Department of Electrical and Computer Engineering,
North Carolina State University, Raleigh, NC 27695-7911}

\author{M. Buongiorno Nardelli}
\affiliation{Department of Physics, North Carolina State University,
Raleigh, NC 27695-8202}
\affiliation{CSMD, Oak Ridge National
Laboratory, Oak Ridge, TN 37831}

\author{K. W. Kim}
\affiliation{Department of Electrical and Computer Engineering,
North Carolina State University, Raleigh, NC 27695-7911}

\begin{abstract}
The electron-phonon interaction in monolayer graphene is
investigated using density functional perturbation theory.  The
results indicate that the electron-phonon interaction strength is of
comparable magnitude for all four in-plane phonon branches and must
be considered simultaneously.  Moreover, the calculated scattering
rates suggest an acoustic phonon contribution that is much weaker
than previously thought, revealing an important role of optical phonons even
at low energies.  Accordingly it is predicted, in good agreement
with a recent measurement, that the intrinsic mobility of graphene
may be more than an order of magnitude larger than the already high values
reported in suspended samples.

\end{abstract}


\maketitle


Graphene, a two-dimensional (2D) sheet of carbon atoms in a
honeycomb lattice, continues to attract much attention due to its
unique physical properties.  Aside from a substantial academic
interest resulting from the relativistic-like behavior of charge
carriers, this material is considered very promising in device
applications as it has an extremely high intrinsic mobility, even at
room temperature.  Although in realistic conditions (i.e., placed on a substrate) the mobility  tends to decrease significantly due to
the presence of additional scattering mechanisms at the interfaces
\cite{Bolotin_PKim_2008,Fuhrer2009,Geim_2008_ripples},
much effort is currently being devoted to eliminate, or at least
minimize, these effects which are detrimental to graphene transport
characteristics.  Therefore, it is crucial to develop an accurate knowledge of
the electron-phonon scattering as it determines the ultimate limit of
any electronic device performance. The strength of electron-phonon coupling is typically estimated
using the deformation potential approximation (DPA); it has been
applied for graphene by a number of authors
\cite{Hwang_Das_Sarma_2008,Vasko,Ferry}. When the corresponding
deformation potential constant was estimated from the transport
measurements, however, the results revealed a discrepancy that is
too large to be ignored
\cite{Bolotin_PKim_2008,Fuhrer2009,hong:136808}.  Moreover, a very
recent observation of mobilities in excess of $10^7$
cm$^2$/V$\cdot$s at $ T \lesssim 50 $~K in the decoupled graphene
\cite{neugebauer:136403} drastically departs from the conventionally
accepted values, raising serious questions about the current
understanding of the intrinsic transport characteristics of
graphene.  A detailed theoretical analysis of electron-phonon
interaction beyond the DPA is clearly called for.

In this work, we apply a first principles approach based on density
functional theory (DFT) to calculate the electron-phonon coupling
strength in graphene.  The obtained electron scattering rates
associated with all phonon modes are analyzed and the intrinsic
resistivity/mobility of monolayer graphene is estimated as a
function of temperature.  The results clearly elucidate the role of
different branches (particularly, the significance of optical
phonons and intervalley scattering via acoustic phonons) as well as
limitations of DPA.  The obtained "effective" deformation potential
constants suggest the possibility that the intrinsic scattering rates
may presently be overestimated, potentially due to the complex
influence of extrinsic mechanisms, including the substrate.
A subsequent Monte Carlo calculation predicts, in good agreement with
Ref.~\onlinecite{neugebauer:136403}, that the mobility of graphene
could be more than an order of magnitude larger than the already
high values reported in suspended samples \cite{Bolotin_PKim_2008}
when presumably all of the extrinsic scattering sources are
eliminated.

A key component of our theoretical study is the evaluation of the
electron-phonon interaction matrix elements from first principles by
employing density functional perturbation theory (DFPT) within the
DFT formalism \cite{Baroni_DFT_review}.  This technique has the
advantage of dealing with an arbitrary electronic state $\mathbf{k}$
and phononic state $\mathbf{q}$ on an equal footing.  Each phonon is
treated as a perturbation of the self-consistent potential created
by all electrons and ions. The calculation of the potential change
due to this perturbation gives the value of the electron-phonon
matrix element:
\begin{equation}
g^{(i,j) \nu}_{\mathbf{k} +
\mathbf{q}, \mathbf{k}} = \sqrt{\frac{\hbar}{2 M \omega_{\nu,\mathbf{q}}}} ~ \langle j,\mathbf{k} +
\mathbf{q} | \Delta V^{\nu}_{\mathbf{q}, SCF} |i, \mathbf{k} \rangle\,,
 \label{g_DFPT}
\end{equation}
where $|i, \mathbf{k} \rangle$ is the Bloch electron eigenstate with
the wavevector $\mathbf{k}$, band index $i$, and energy $E_{i,
\mathbf{k}}$; $\Delta V^{\nu}_{\mathbf{q}, SCF}$ is the derivative
of the self-consistent Kohn-Sham potential \cite{Baroni_DFT_review}
with respect to atomic displacement associated with the phonon from
the branch $\nu$ with the wavevector $\mathbf{q}$ and frequency
$\omega_{\nu, \mathbf{q}}$, and M is the atomic mass.  A similar
calculation has previously been performed in graphene for
$\mathbf{q}$ at high symmetry points $K$ and $\Gamma$, where Kohn
anomalies occur (see below) \cite{Piscanec_2004,Lazzeri_2008}.
However the scattering rate calculation requires summation over the
entire first Brillouin zone (FBZ). Using DFT and DFPT (Quantum Espresso software
\cite{Q_E}), the electronic band structure and the phonon dispersion
are calculated for a single graphene sheet in vacuum. The
computation is performed on a hexagonal unit cell with the two atom
basis, lattice constant $a$ of 2.42 \AA\, and approximately 5 \AA\ of
vacuum space, based on a norm-conserving pseudopotential and the generalized gradient approximation/local density approximation
\cite{Marzari_2005}. A Monkhorst-Pack grid \cite{Monkhorst} of
36$\times$36$\times$1 is used with no offset for integration.  To
determine the scattering rate of an electron due to the interaction
with the phonons, the matrix elements $g^{(i,j) \nu}_{\mathbf{k} +
\mathbf{q}, \mathbf{k}} $ are extracted from the self-consistent
electron-phonon calculation on the Monkhorst-Pack grid of
$\mathbf{q}$-points that covers the entire FBZ and incorporates the
points of high symmetry: $\Gamma$, $K$, and $M$.

\begin{center}
\begin{figure}[tbp]
\includegraphics[bb=40 133 713 540,width= 8 cm]{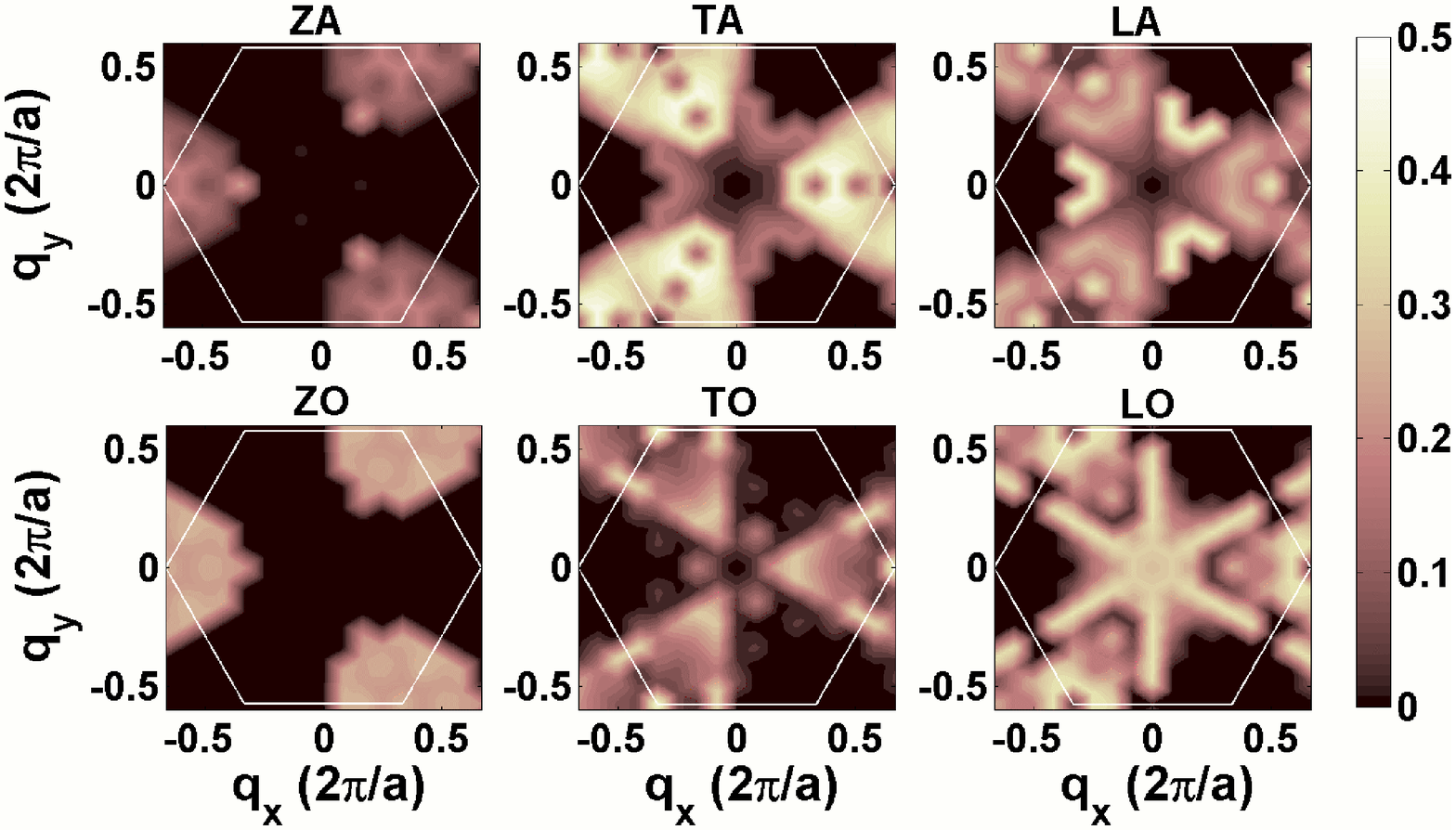}
\caption{(Color online) Electron-phonon interaction matrix elements
$\left| g^{(i,j) \nu}_{\mathbf{k} + \mathbf{q}, \mathbf{k}} \right|$
(in units of eV) calculated by DFPT for $\mathbf{k}$ at the
conduction band minimum (i.e., the Dirac point) as a function of
phonon wavevector $\mathbf{q}$.  High anisotropy beyond the long
wavelength approximation is clearly visible. The presence of Kohn
anomalies reported earlier is revealed as the peaks (light color) at
three equivalent Dirac points for TO and at the center of the FBZ
for LO.}  \label{FIG_g_DFPT}
\end{figure}
\end{center}

Figure~\ref{FIG_g_DFPT} shows the resulting matrix elements for an
electron at the Dirac point [i.e., $\mathbf{k} = (4\pi/3a,0)$] as a
function of the phonon wavevector $\mathbf{q}$.  The contribution of all
six phonon branches is considered, whereas the electronic states
(both initial and final) are limited to the lowest conduction band.
The matrix elements display a pronounced anisotropy everywhere
except near the FBZ center ($\mathbf{q} = 0$), where they are in
qualitative agreement with the DPA (i.e., nearly isotropic) and also
comply with the group theory analysis of electron-phonon interaction
in graphene (in-plane modes) \cite{Manes2007}: lim$_{q \rightarrow
0}|g^{ac}|\sim q$ and lim$_{q \rightarrow 0}|g^{op}|\sim constant$
for acoustic and optical phonons, respectively.  Another important
feature is the Kohn anomaly $-$ a strong coupling of the electrons
with in-plane optical phonon branches in points of high symmetry
\cite{Piscanec_2004}.  This is prominently illustrated by the three sharp peaks at three equivalent $K$ points in the TO mode. The other Kohn anomaly (LO
branch, $\mathbf{q}$ at the $ \Gamma$ point) is not as distinct.
Even more crucial is that their magnitude can be directly validated
against experimental measurements. Comparison with available experiments and
another \emph{ab initio} approach (GW) clearly shows good agreement,
though DFT is known to underestimate the electron exchange
correlation energy in the presence of Kohn anomaly. Specifically,
the GW calculation and the experiments (Raman spectroscopy and
inelastic x-ray scattering - see Ref.~\onlinecite{Lazzeri_2008} and
references therein) provide values for $|g|^2$ that are larger
than our results by about 15 \% and 60 \% for $\mathbf{q}$ at the
$\Gamma$ (LO) and $K$ (TO) points, respectively. This discrepancy,
however, does not lead to a substantial difference in the calculated
scattering rates, since the correction is appreciable only in a very
small area of the FBZ close to the symmetry points. Accordingly, the
scattering matrix elements obtained by DFPT are expected to provide
sufficient accuracy to calculate the intrinsic transport properties
of monolayer graphene.

\begin{center}
\begin{figure}[tbp]
\includegraphics[bb=14 13 326 245,width=8cm]{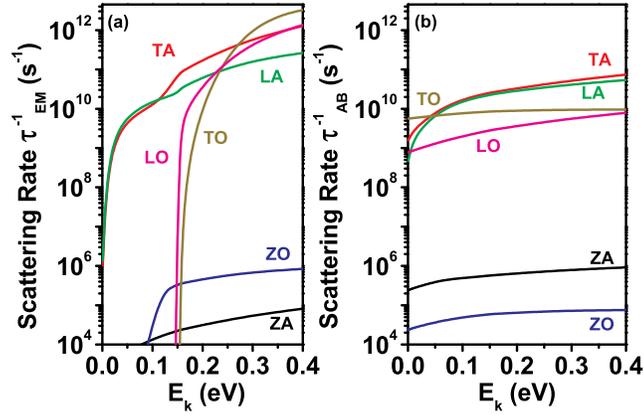}
\caption{(Color online) Phonon (a) emission and (b) absorption
scattering rates at $T = 300 $ K as a function of electron energy
$E_\mathbf{k}$, as $\mathbf{k}$ changes along the $K$-$\Gamma$
direction in monolayer graphene. The contributions of all six
branches are shown separately. The out-of-plane phonons (ZA and ZO)
do not play an important role.  At very low electron energies ($E <
k_{B}T $), the scattering with absorption of TO phonons is
dominant.}  \label{FIG_R_DFPT}
\end{figure}
\end{center}

Assuming low electron concentrations in the conduction band, the
electron-phonon scattering rate can be readily obtained by using the
Fermi's golden rule.  Figure~\ref{FIG_R_DFPT} shows the results of the calculation at room
temperature ($T = 300$ K) based on the DFPT matrix elements
described above [Eq.~(\ref{g_DFPT})].  Although the result is
plotted specifically for electrons with wavevector $\mathbf{k}$
along the $K$-$\Gamma$ direction, its directional dependence is
minor in the energy range under consideration.  The total electron
scattering rate (from all six branches) for $E \sim k_{B}T$ is of
the order of $10^{10}$ s$^{-1}$.  This is lower than one might
expect based on earlier estimates
\cite{Fuhrer2009,Hwang_Das_Sarma_2008,Ferry}; a discussion on the
possible reasons is given below.  The scattering rates for the
out-of-plane phonons (ZA and ZO) are much smaller (at least three
orders of magnitude) in comparison with the rest of the branches
(in-plane modes).  In 2D structures with an in-plane reflection
symmetry (e.g. graphene), only in-plane phonon modes can couple
linearly to electrons \cite{Manes2007}.  On the other hand, all four
in-plane branches contribute comparable scattering rates at room
temperature despite the differences in phonon dispersion.
Accordingly, optical phonons can play an important role even at low
electron energies. Another interesting feature is a change of slope
in the emission rates of TA and LA phonons
[Fig.~\ref{FIG_R_DFPT}(a)]. This occurs due to the onset of
zone-edge phonon emission near the $K$ points (denoted as
$\mathbf{q}= K$ for simplicity) leading to intervalley transfer of
electrons. As shown, intervalley scattering by TA phonons can be
important since the emission threshold is relatively low: $\hbar
\omega_{K}^{TA} \simeq 120$ meV is substantially smaller than the
corresponding TO and LO phonon energies.

\begin{center}
\begin{figure}[tbp]
\includegraphics[bb=14 9 210 171,width=8cm]{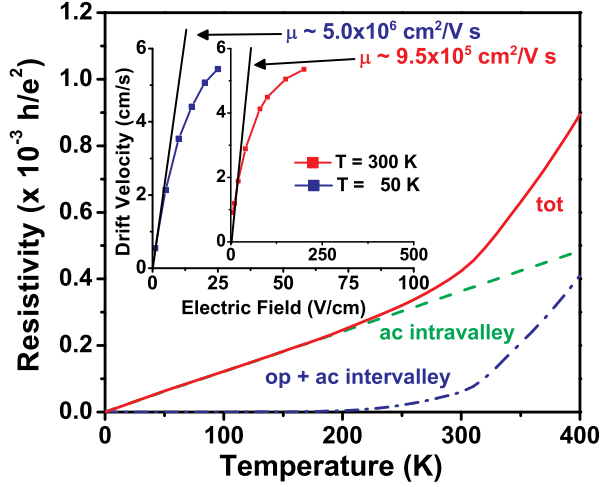}
\caption{(Color online) Intrinsic resistivity as a function of
temperature. At $T<200$ K, the quasielastic intravalley scattering
with in-plane acoustic phonons is dominant (dashed line).  At higher
temperatures, the contributions of both optical phonons and
intervalley scattering by TA and LA modes (dashed-dotted line) lead
to an exponential growth of the resistivity. The inset shows the
low-field mobilities at $T= 50$ and 300 K, as determined from the
slope of the drift-velocity versus electric field characteristic
obtained by a full band Monte Carlo simulation. }
 \label{FIG_rho_vs_T}
\end{figure}
\end{center}

The obtained scattering rates can be used to predict the intrinsic
transport properties of graphene.  The temperature dependence of the
electrical resistivity is of particular interest as it is believed
to be relatively insensitive to the impact of environment (e.g.,
substrate) \cite{Fuhrer2009}, potentially enabling direct comparison
with measurements in real (non-ideal) samples.  Following the
relaxation time approximation, an estimate of the intrinsic
resistivity can be expressed straightforwardly as:
\begin{equation}
\rho_{i} \approx
  \left[ e^2 D(E_F)\frac{v_F}{2}
\tau_{tot}(E_F) \right]^{-1} \,,
 \label{rho_RTA}
\end{equation}
where $D(E_F)$ is the electronic density of states at the Fermi
level $E_F$, $\tau_{tot} (E_F)$ denotes the total scattering rate at
$E_F$, and $v_F = 10^8$ cm/s is the Fermi velocity.
Figure~\ref{FIG_rho_vs_T} shows the result of this calculation for
$n = 10^{12}$ cm$^{-2}$. At relatively low temperatures ($T < $ 200
K), the electrons mostly scatter quasielastically with TA and LA
phonons (intravalley scattering), leading to a linear temperature
dependance of resistivity. As the temperature increases ($T > 200$
K), the contributions of both optical phonons and intervalley
scattering by TA/LA modes increase. The exponential slope in this
region signifies the sensitivity of large-energy phonon occupancy to
temperature.

Within the DPA, the slope of the intrinsic resistivity in the linear
region is proportional to the square of the acoustic phonon
deformation potential constant \cite{Hwang_Das_Sarma_2008}. Using
the DFPT based results, we estimate a value of the "effective"
deformation potential constant to be $ D_{ac} \simeq 4.5 $ eV for
the combined contribution of TA and LA phonons.  When compared to
those extracted from the experiments, this number is fairly close to
the recently reported 7.8 eV \cite{hong:136808}, while much smaller
than others (29 eV \cite{Bolotin_PKim_2008}, 18 eV
\cite{Fuhrer2009}). On the theory side, however, $D_{ac}$ of similar
magnitude (2.6 eV) was also deduced from a valence force model
\cite{Perebeinos2009}.  With the estimates under non-intrinsic
conditions (i.e., experiments) consistently larger than the
intrinsic theoretical prediction, one likely explanation is that the
influence of the substrate on graphene electron transport may
currently be underestimated.  This is the most apparent interpretation
of the wide spread between experimental data (7.8$-$29 eV)
indicative of additional factors or mechanisms in play.

The issue of identifying intrinsic scattering characteristics is
further examined by comparing our calculations with the results in
the decoupled graphene layers (allegedly, the purest form of
graphene), where the electron mobilities in excess of $10^7$
cm$^2$/V$\cdot$s were obtained from measurements at low temperatures
$T \lesssim 50$ K \cite{neugebauer:136403}. For this, a full-band Monte
Carlo simulation is performed.  The model utilizes the complete
phonon spectrum and electron-phonon scattering rates as determined
by DFT/DFPT, described above. The inset of Fig.~\ref{FIG_rho_vs_T}
shows the obtained velocity-field curves, where the low-field
mobility is estimated from the slope in the linear region. The
calculation produces a very high number of approximately $ 5 \times
10^6$ cm$^2$/V$\cdot$s at 50 K, in good agreement with
Ref.~\onlinecite{neugebauer:136403}.  Accordingly, the results
clearly indicate the accuracy of the DFPT scattering rates,
particularly those by TA and LA  phonons, as the low-field mobility
in this temperature range is dominated by the interactions with
acoustic modes (see Fig.~\ref{FIG_rho_vs_T}).  Note that the
calculated strength of the electron$-$optical phonon interaction was
validated earlier in connection with the Kohn anomaly.  A corresponding estimation at 300 K predicts the intrinsic mobility approaching $ 10^6$ cm$^2$/V$\cdot$s that is also much higher than previously thought.

\begin{center}
\begin{figure}[tbp]
\includegraphics[bb=15 15 281 234,width=7cm]{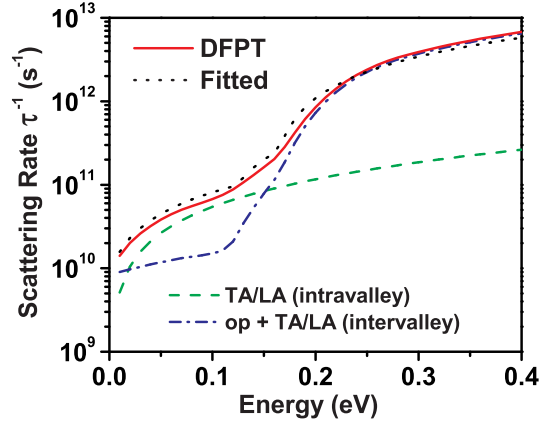}
\caption{(Color online) Total scattering rate  as a function of
electron energy at $T = 300$ K. The dotted line shows the fitted
curve calculated from Eq.~(\ref{R_fitting}).  Clearly, the role of
optical phonon scattering and intervalley transfer must be taken
into account in most cases judging from the discrepancy between the
solid (total) and dashed (TA \& LA intravalley) lines.}
\label{FIG_R_DFPT_and_fitting}
\end{figure}
\end{center}

For practical application, it would be convenient to approximate the
\emph{ab initio} scattering rates by a simple analytical model.
Presently, the expression commonly used for the acoustic phonon
scattering is given as  \cite{Hwang_Das_Sarma_2008}
\begin{equation}
\left(\frac{1}{\tau_{\mathbf{k}}}\right)_{ac} = \left(\frac{k_B}{4 \hbar^3
v_{F}^2 \rho_{m}v_{s}^2}\right) D_{ac}^2 T E_{\mathbf{k}} \,,
 \label{R_LA_Das_Sarma}
\end{equation}
while
\begin{equation}
\left(\frac{1}{\tau_{\mathbf{k}}}\right)_{op} =
\frac{D_{0}^2}{\rho_{m}\omega_{0} (\hbar
v_{F})^2}~\left[ \left(E_{\mathbf{k}} - \hbar
\omega_{0}\right)\left(N_{\mathbf{q}}+1\right)
\Theta\left(E_{\mathbf{k}} - \hbar \omega_{0}\right) +
\left(E_{\mathbf{k}} + \hbar \omega_{0}\right)N_{\mathbf{q}} \right]
 \label{R_opt_Ferry}
\end{equation}
is adopted for the optical phonon scattering \cite{Ferry}.  Here, $v_s$
denotes the sound velocity, $\rho_m$ the mass density,
$N_{\mathbf{q}}$ the phonon occupation number, and $\Theta(x)$ the
Heaviside step function.  Although these formulae are valid under
limited conditions [for example, Eq.~(\ref{R_LA_Das_Sarma}) takes
into account only the LA phonons in the long wave approximation],
the deformation potential constants $D_{ac}$ and $D_{0}$ can be
treated as effective quantities to determine the contribution of all
the relevant modes. A similar discussion was made above in
reference to the slope of the electrical resistivity. Then, the
total rate can be represented as a sum:
\begin{equation}
\left( 1/\tau \right)_{tot} = \left(1/\tau\right)_{ac,KK} +
\left(1/\tau\right)_{ac,KK'} + \left(1/\tau\right)_{op}  \,.
\label{R_fitting}
\end{equation}
Of the combined contribution of TA and LA phonons, the first term
$\left(1/\tau\right)_{ac,KK}$ represents electron intravalley
scattering described by Eq.~(\ref{R_LA_Das_Sarma}), whereas the
intervalley transfer $\left(1/\tau\right)_{ac,KK'}$ is calculated by
Eq.~(\ref{R_opt_Ferry}), as there is no distinction between acoustic
and optical modes near the zone edge.  The last term accounts for
the interactions with both TO and LO modes.
Figure~\ref{FIG_R_DFPT_and_fitting} shows the results with the
following parameters at 300 K: $\left(1/\tau\right)_{ac,KK}$ $-$
$D_{ac} = 4.5 $ eV; $\left(1/\tau\right)_{ac,KK'}$ $-$ $ D_{0} =
3.5\times10^{8}$ eV/cm, $ \hbar \omega_{0} = \hbar \omega_{K}^{TA} =
124$ meV; and  $\left(1/\tau\right)_{op}$ $-$ $D_{0} =1\times10^{9}$
eV/cm, $ \hbar \omega_{0} = \hbar \omega_{K}^{TO} = 164.6$ meV.
Clearly, Eq.~(\ref{R_fitting}) not only serves as an excellent
analytical approximation but also reveals the interplay among
different phonon scattering processes at a given temperature and
electron energy.  This simple model is expected to remain valid for
$T > T_{BG}$ with the Bloch-Gr\"{u}neisen temperature $T_{BG}
\approx 50-60$ K \cite{Hwang_Das_Sarma_2008}.


In conclusion, our first principles analysis clearly illustrates
that all in-plane phonons play an important role in electron-phonon
interactions in graphene and must be considered for transport
studies at room temperature.  Moreover, the results suggest that the
influence of the substrate may be more significant than previously
understood, clarifying at least in part the discrepancies observed
in the strength of electron-acoustic phonon scattering. Under ideal
conditions, it is predicted that the mobility of graphene could
reach as high as $10^{6}$ cm$^2$/V$\cdot$s at room temperature.

This work was supported, in part, by the DARPA/HRL CERA, the FCRP
FENA programs and NSF-ECS0621776. MBN wishes to acknowledge partial
support from the Office of Basic Energy Sciences, U.S. Department of
Energy at Oak Ridge National Laboratory under contract
DE-AC05-00OR22725 with UT-Battelle, LLC.  MBN and JTM are thankful to
M. Lazzeri for useful discussions.

\newpage

\newpage

\newpage

\newpage

\end{document}